# Neural Networks, Artificial Intelligence and the Computational Brain

Martin C. Nwadiugwu

[1] University of Ilorin, Nigeria.

*Abstract: In recent years, several studies have provided insight on the functioning of the brain which consists of neurons and form networks via interconnection among them by synapses. Neural networks are formed by interconnected systems of neurons, and are of two types, namely, the Artificial Neural Network (ANNs) and Biological Neural Network (interconnected nerve cells). The ANNs are computationally influenced by human neurons and are used in modelling neural systems. The reasoning foundations of ANNs have been useful in anomaly detection, in areas of medicine such as instant physician, electronic noses, pattern recognition, and modelling biological systems. Advancing research in artificial intelligence using the architecture of the human brain seeks to model systems by studying the brain rather than looking to technology for brain models. This study explores the concept of ANNs as a simulator of the biological neuron, and its area of applications. It also explores why brain-like intelligence is needed and how it differs from computational framework by comparing neural networks to contemporary computers and their modern day implementation.*

## 1. BACKGROUND

The human brain has been undergoing serious investigations by many researchers in the field of neuroscience. In time past, there have been considerable investigations of the structure of the brain (anatomy of the brain), but studies on the functional operation of its complex neural network paraded all sorts of fantasies as knowledge for many centuries (Sundal *et al*., 2014). Around the middle of the 18th century, a functional understanding of the human brain began to take shape. During this period, studies revealed that impulse of nerve previously considered as "animal spirits" are simply electric signals similar to a charge in an electrical circuit (Sundal *et al*., 2014). Moreover, improvements in neuroscience research and microscopy have showed the structure and form of neurons, representing the brain as a network of neurons that communicate using chemical signals.

The adult brain of a human consists of about 100 billion neurons, each with about 1,000 - 10,000 connections, making a total $10^{14}$-$10^{15}$ connections in the brain (Sundal *et al*., 2014). It is perhaps the most complex system, more complex than the entire mobile network of the world, with its neurons making and unmaking connections at a timescale that can be as short as a few tens of seconds (Sundal *et al*., 2014). How thoughts are being processed by the human brain has remained an enigma up till today. Significant ventures in the field of artificial intelligence (AI) have enabled scientists to come close to the nature of thought processes inside a brain (Zhang, 2011). In the area of AI, Artificial Neural Networks (ANNs) is being employed in computational tools to model a biological brain (Willamette, 2014). AI seeks to answer questions like "how network of neurons in the visual processing areas of the brain transduce the optical image that falls on the retina and how they can be simulated to make intelligent device"; however, answers to these questions are best described in the language of mathematics, which is the primary preoccupation of computational neuroscience (Sundal *et al*., 2014).

Computation for brain-like systems involves comprehending the biological brain using computational models (computational neurobiology) and the process of building and animating a machine to emulate the biological brain (neural computing). All these and other diverse neural network specialties are examined in the emerging field of computational neuroscience. The overarching goal of the study was to 1) understand how artificial neural networks are modelled and how they differ from their biological counterparts 2) their application and implementation particularly in the field of medicine. The remaining sections of this paper will highlight the structure and composition of biological and artificial neural networks, the purpose of the study and the methods used. An overview of the historical background and architecture of ANNs and their application particularly in neuroscience and medicine will be discussed in relation to AI. Next, the similarities and contrast between the brain and the computer and the rationale for brain-like intelligence will be explored. The review will conclude with a recommendation on how further studies on the use of ANNs will be beneficial.

## 2.0  ANATOMY OF A NEURON

The foundational unit of the human brain is the neuron which refers to the nerve cell and its processes (Snell *et al.*, 2010). Neurons are found in ganglia of the brain and in the spinal cord (Snell *et al.*, 2010). They receive input from other neurons and are specialized cells capable of reacting to stimuli and conducting nerve impulses (Snell *et al.*, 2010). Neurons differ in shape and size, and each neuron has a cell body consisting of processes called neurites (Snell *et al.*, 2010). The neurites also known as dendrites, are responsible for receiving and conducting information towards the cell body (Snell *et al.*, 2010). The term nerve fiber is used to refer to the dendrites and axons which conduct impulses away from the cell body as can be seen in Fig.1.0 (Snell *et al.*, 2010).

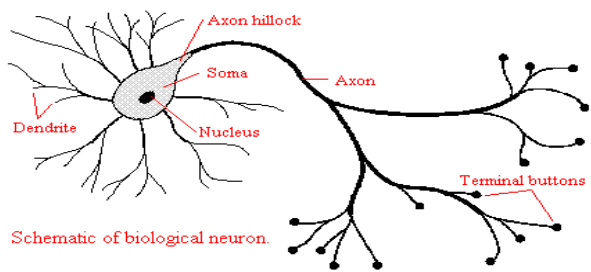

Fig. 1.0 Structure of a Neuron (Papadourakis, 2014)

Figure 1.0 shows a biological neuron with various parts which includes:

(a)  The Dendrites
(b)  The Cell body or Soma
(c)  The Axon

The neurons have branched projections called dendrites which helps to direct electrochemical stimulation received from neighboring neural cells via synapses located at a number of points around the dendritic arbor to the cell body of the neuron from which the dendrites project (Mujeeb, 2012). Dendrites are important in uniting synaptic inputs and deciding the degree neurons produce action potentials (Snell *et al.*, 2010). It has been reported that dendrites release neurotransmitters and support action potentials which were previously thought to be axon specific (Mujeeb, 2012). The soma is where signals from the dendrites are joined and passed on; they serve to maintain the cell and keep the neuron functional but they do not play an active role in the transmission of neural signal (Snell *et al.*, 2010). The axon is a slender, long protruding part of a neuron that directs impulses away from the soma; they differ from dendrites in function, length and shape (Mujeeb, 2012). Neuron do not have more than one axon, as most axons sometimes branch extravagantly, those with no axon transmits signals from their dendrites (Snell *et al.*, 2010; Mujeeb, 2012).

## Biological and Artificial Neurons

The biological neurons constitute the foundational blocks of the human brain. Artificial neurons on the other hand forms ANNs which are computer algorithms inspired by biological neuron and modeled after the brain to perform specific computational tasks (Mano, 2014).

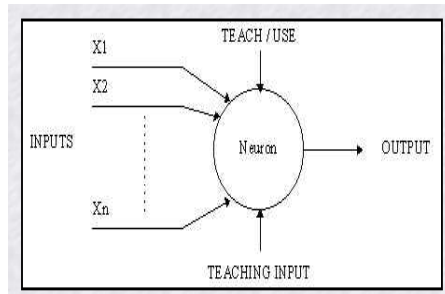

Fig. 1.1 Artificial Neuron Model (Stergiou and Siganos, 2014)
.

The above figure 1.1 shows a model of an artificial neuron described in 1943 by McCulloch and Pitts. It consists of inputs with an electrical impulse represented as (1) or with no electrical impulse represented as (0) (Reed 2015). A weight is associated with each input value which is multiplied by the activation function. If the weighted sum of the input is greater than or equal to $\theta$, then the neuron fires and returns 1, if not, it does not fire but returns 0 (Reed 2015). In addition, a neuron has a threshold for activation and a number of weighted associations to neighboring neurons (Mano, 2014). When the sum of the activation a neuron receives from its neighbors is greater than its threshold of activation, it fires and pass along this intensity to neighboring neurons. Training the network involves modifying associated weights of connections to perform certain task which accounts for learning (Mano, 2014). Artificial neuron forms the basis of artificial neural networks (ANNs). In the 1950's, it became a focus of computer science research because at that time, it was said that humans do not have the power of remembering and the speed of computers but are capable of intricate actions and reasoning. Interconnecting neurons in general form networks which are of two main types, namely, ANN and biological neural network (BNN).

## 3.0  NEURAL NETWORKS

### Biological Neural Networks

The neural network in the brain is a mutually joined web of biological neurons conducting complex distributions of electrical signals. The human brain can anatomically be distinguished into several divisions such as the cortex, brainstem, cerebellum etcetera. It can further be subdivided into several areas and regions according to the functions performed by them, and the in-built structure of the neural network anatomy (Mano, 2014).

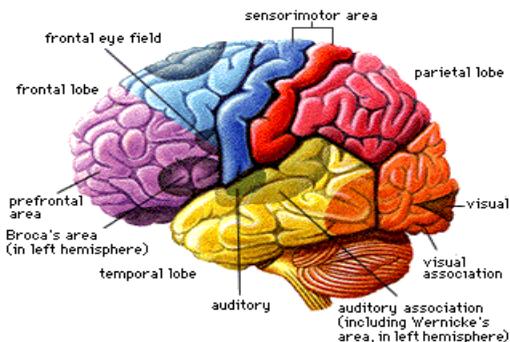

Fig. 2.0 Anatomical areas of the Human Brain (Willamette, 2014)

Figure 2.0 shows different anatomical areas of the human brain consisting of biologically interconnected neurons. In biological neural network dendrites receive input signals which fires an output based on the input signals (Dharani, 2015). The overall pattern of neural linkage (projections) between the various areas is very complicated, and not all information about them is known.

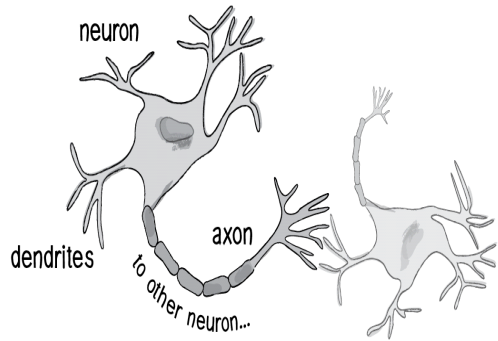

Fig. 2.1 A Biological Neural Network (Shiffman, 2014)

The above figure 2.1 shows a biological neural network formed by an interconnected nerve cell. Apart from forming long-range connections with neighboring neurons, they also link up with other nerve cells to form complex and dense local networks. The general brain architecture contains many networks of interconnected neurons, each utilizing biochemical reactions to send, receive and process information. The synapse is a small gap where the terminal button of each neuron is connected to other neurons which can be seen in Fig 2.2 which shows a neuron forming a synapse through the connection of its axon endings with the dendrites of another neuron. The dendrite serves as the input device which receives electrical signals or impulses from other neurons. The cell body or soma gives a summation of inputs from the dendrites which causes excitation or inhibition. When the summation of inputs from the dendrites is greater than a certain threshold, the neurons are excited and fires along the axon resulting in an output (Mujeeb, 2012).

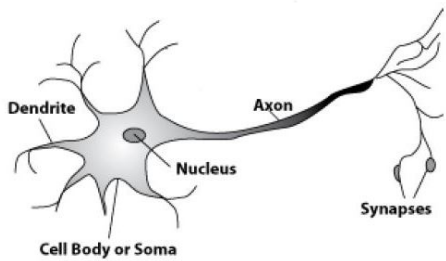

Fig. 2.2 A neuron forming synapse (Mujeeb, 2012)

A neuron typically receives input from other neurons via dendrites as seen in figure 2.3. The dendrites are connected to many neighboring neurons in a way that a negative or positive charge is received by a dendrite when a neurons fires (Mann 2011). The strengths of all charges received are added approximately together via temporal and spatial summation process, and once the summed input goes beyond a sufficient level, an electrical pulse is discharged which travels to the next neuron(s) or receptor(s) through the axon (Mann 2011). This activity results in depolarization and a period of refraction where the neuron is less excited and unable to fire. The final part of the axon of a neuron form the output zones that just about touch the dendrites of the next neuron. Chemicals called neurotransmitters are released from a neuron and they effect the transmission of impulses to the next neuron binding at their receptor to form a linkage between the two neurons called a synapse (Fig.2.3) (Mujeeb, 2012).

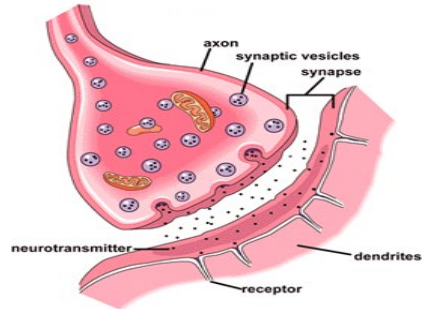

Fig. 2.3 Synapse (Tewari, 2012)

The word "synapse" comes from the term "synaptein" which Sir Charles Scott Sherrington and colleagues coined from the Greek word "syn-" which means together and "haptein" which means to clasp (Cooper nd.). Synapses are vital to neuronal function and are the means by which neurons pass signals to individual target cells (Cooper nd.). At the synapse, the plasma membrane of the presynaptic neuron comes into close juxtaposition with the postsynaptic cell membrane (Cooper nd.). Both pre- and postsynaptic locales contain a collection of molecular system that carry out the signaling process and link their membranes together (Mujeeb, 2012). The presynaptic part is located on an axon in many synapses, while some presynaptic sites can be found on a dendrite or cell body (Mann 2011). Two different basic type of synapses exist, namely, (a) The chemical synapse and (b) The electrical synapse.

For chemical synapse, "the presynaptic neuron releases a chemical called a neurotransmitter that binds to receptors located in the postsynaptic cell, usually embedded in the plasma membrane" (Mujeeb, 2012). The neurotransmitter sometimes initiates an impulse or molecular relaying receptor signals that could inhibit or excite post-synaptic neurons (Dharani, 2015). In an electrical synapse, electrical current can pass through the postsynaptic and presynaptic cell membranes which are connected by gap junctions causing voltage changes in the postsynaptic cell induced by presynaptic cell voltage changes (Mujeeb, 2012). Electrical synapses rapidly transfer signals from one cell to the next, which is a major advantage. The degree to which signals are passed from one neuron to the next depends on the amount of available neurotransmitter, the receptor arrangement, number of receptor and the amount of reabsorbed neurotransmitter etcetera (Mujeeb, 2012).

### Artificial Neural Networks (ANN)

ANN is a computational system that processes information in parallel collectively in every part of a network of nodes (neuron) (Shiffman, 2014). In ANNs the individual neurons (also called nodes) receive input, process and produce an output; the system of closely connected numerous neurons exhibits intelligent and rich behaviors (Shiffman, 2014). The ANN can be described as a data processing pattern influenced by manner the human nervous systems process information (Stergiou and Siganos, 2014). Analysis of ANNs show that it is a method of data analysis, which imitates the method and functioning of the human brain. It consists of many eminently joined neurons working in accord towards a specific task (Rhouma et al. 2011). As part of AI, ANNs tries to make computers mimic the capabilities of human brain in the way it processes information (Willamette, 2014). The ANNs are programmed to replicate the functions of the neural structure/architecture of the human brain. They are adaptive, flexible, adjusting and learning with numerous and diverse external and internal stimulus (Willamette, 2014).

The capabilities of ANNs has been deployed over the years in problems with varying degrees of complexity and in diverse fields of application that depicts how neural networks probably function and learn in a biological environment (Tayarani et al. 2013). The computational characteristics of ANNs include models designed with an ability to organize, learn, adapt, or generalize data. The process of learning occurs by training with examples using algorithms to iteratively adjust the weights of interconnected neurons to bring about the desired input–output relationships (Kustrin and Beresford 2000). This process has been used to a great extent in modelling, optimization, calibration, and pattern recognition. The ANNs are applied in medicine for the diagnosis of diseases, and have shown good prospects in calculating the biological and physio-chemical properties of drugs. In recent years, the pharmaceutical applications of ANN reviewed by Kustrin and Beresford (2000) showed that the aqueous solubility of drugs using several molecular descriptors have been calculated using ANNs. It has been suggested that by designing, testing and using the appropriate ANNs, predictions of the binding energy of drugs on basis of structural descriptors describing the structure of selected basic drugs could be possible (Tayarani et al. 2013).

## Historical Background of ANNs

Although ANNs imitative representations seem to be recently advancing, this area of work was historically established before the advent of computers (Stergiou and Siganos, 2014). The logician Walter Pitts and the neurophysiologist Warren McCulloch produced the first ANNs in 1943 (Stergiou and Siganos, 2014), but could not achieve much at that time as a result of the technology available. The use of inexpensive computer emulations enabled important advances in neural network simulations (Rhouma et al. 2011). There were periods of excitement initially, followed by periods of frustration when professional and monetary support was minimal, as further vital development in the field were made a few researchers (Stergiou and Siganos, 2014). *Minsky* and *Papert* published a book in 1969 summing up this obstacles and frustrations among researchers. However, other pioneers of neural network simulations thereafter developed effective technology that went beyond the drawbacks noted by *Minsky* and *Papert* (1969) and were accepted by many without further analysis (Stergiou and Siganos, 2014). At this time with advances in AI and machine learning, the field takes pleasure in the renewal of interest and increased funding.

### *General architecture of Artificial Neural Network (ANN)*

Artificial neurons are the basic units of ANNs; they are interconnected with no central control and work in parallel (Rhouma et al. 2011). Neurons in ANNs are often organized into layers, with one layer connected to adjacent layers. Each neuronal interconnection has an associated weight and updating these weights leads to learning within the network (Mano, 2014). The ANNs receives inputs, and produces outputs depending on weight and the way the interconnections change the input signals which undergoes modification and is transmitted to connected neurons if the received signal exceeds a certain threshold (Czogala and Leski 2012); the resulting output is generated by the non-propagating output layers (layers that do not produce signals) (Mano, 2014). Neurons in an ANNs are sequentially structured into layers and linked to neighbors by variable connection weights, and each layer can have several different neurons with various transfer functions (Czogala and Leski 2012). ANNs learns by example just like the human brain which learns by varying the connection strengths between neurons which involves removing or summing connections between neurons; stronger connections are frequently activated (Willamette, 2014; Owens & Tanner 2017). ANNs are configured for a particular application (e.g. data classification, pattern recognition) through a learning process similar with biological systems that involves adjusting synaptic connections existing between the neurons. The composition of ANNs are often categorized into layers of communication such as: 1) Input layers 2) Hidden and 3) Output layers. The input layer is configured to a set of values, while computations by the hidden layers update the values from the input layer to the output layers (Russel *et al*, 2002). The variety of tasks where ANNs have been applied includes:

**(a) Artificial Intelligence**: ANN can be applied in estimating mathematical functions and extracting image features for optical character recognition. The autonomous land vehicle an ANN was used to retrieve features about the road to navigate an unmanned vehicle by Carnegie Mellon University's NAVLAB (Luebbers & Pandya 1994). They have been incorporated in game playing, voice recognition, detection of explosives and filtering unsolicited emails (Federici et al. 2007)

**(b) Learning**: This can be supervised or unsupervised learning. Backpropagation is the most popular method for training the networks, and it is simply a statistical method that update weights based on its difference from the desired output, and various algorithms can be used to search for the optimal set of weights (Heidelberg, 2005). Gradient decent, an optimization method that searches at each step in the direction that comes closest to the goal is the most common algorithm (Heidelberg, 2005; Patel et al. 2014).

**(c) Adaptive learning**: An ability to learn how to do tasks based on the data given for training or initial experience.

**(d) Self-Organization**: An ANN can create its own organization or representation of the data it obtains during learning period.

**(e) Fault Tolerance via Redundant Information Coding:** the ability of network to retain its capabilities following major network damage (Heidelberg, 2005; Patel et al. 2014).

**(f) Real Time Operation**: these involves computations undertaken in parallel using special physical devices (Heidelberg, 2005).

## Reasons for the study of ANNs in Neuroscience and Medicine

ANNs are much simpler than their biological counterparts as they are composed of fewer artificial neurons than the approximately 100 billion neurons that make up the human brain (Poole, 2010). ANNs are of great interest in neuroscience because they help model and understand aspects of biological neural systems and behavior by simulating the neural systems of simple animals (Poole, 2010). It has been hypothesized that a unique way of understanding the functions of the brain and building intelligence is by using models of the brain (Catalay 2014). Since the brain inspires a model of computation that differs with how computers are being programmed, ANNs becomes an interesting area of study that seeks to implement this paradigm. However, unlike ANNs the brain consists of asynchronously distributed concurrent processes, with no master controller, unlike modern day computers that have inert memory and processors (Poole, 2010).

ANNs is an area of research interest in medicine and it is believed that in the coming years there would be more relevant application to biomedical systems. Currently, research in this area is focused on recognizing diseases via scans for example, the use of CAT scans, cardiograms, ultrasonic scans, modeling systems and parts of human body etcetera (Stergiou and Siganos, 2014). The recognition of diseases using scans and ANNs reduces the need for a specific disease identifying algorithm and how to recognize the disease since the network learns by training from a set of selected example that represents the variations of the disease of interest. In the field of medicine, ANN can be applied in the following:

(a) **Diagnosing biological System via Neural Modeling:** Neural modelling is used in medical diagnosis to model human biological systems in such a way that by comparing the model with real time physiological measurements of vital signs at different physical activities adverse health conditions could be diagnosed quite early making disease detection much easier (Stergiou and Siganos, 2014). If this model is tailored to an individual it very well becomes a model representing that individual's physical condition and can undergo adaptation without an expert supervision (Stergiou and Siganos, 2014). Sensor fusion, an aggregation of various sensor values in order to learn complicated associations among sensor values is another application of ANN. Sensor fusion is useful in medical diagnosis and modelling as it allows for the detection of complicated health conditions by aggregating data from several sensors (Stergiou and Siganos, 2014).

(b) **Electronic noses:** This is an area of ANNs application in telemedicine at remote surgical locations to detect odors which can be transmitted by means of a computer to be recreated at a location with an odor generation system. Telemedicine are medical activities done through communication links over long distances. This transmission is essential to a surgeon because tele-smell enhances tele-present surgery (Stergiou and Siganos, 2014).

(c) **Instant Physician**: This application was developed in the mid-1980s; it trained an auto-associative memory neural network to archive medical records for individual patient (Pomi and Olivera 2006). After training, the application (Instant physician) can be presented with an input in order to find the correct pattern that represents the optimal treatment and diagnosis (Stergiou and Siganos, 2014). The ability of a neural network to adjust its structure and learn over time is what makes it quite essential. ANNs have other diverse application such as:

(1) Pattern Recognition: common application of artificial neural networks used in facial recognition, optical character recognition, etcetera (Shiffman, 2014).
(2) Time Series Prediction: to forecast changes in stock market prices, whether predictions.
(3) Signal Processing: ANNs can be applied to amplify important sounds and filter out noise in cochlear implants and hearing aids, and may be taught to appropriately filter an audio signal and process them (Shiffman, 2014).
(4) Control: self-driving cars employ ANNs to simulate and manage steering decisions (Shiffman, 2014).
(5) Soft Sensors: This refers to the process of studying an array of measurements closely. The ANNs can be used to process and evaluate data input from numerous sensors (NRC, 2005), for example, a thermometer can give us information about the temperature of the air, however, with ANNs we can get additional information on humidity, barometric pressure, dew point, air quality, air density, etcetera.
(6) Anomaly Detection: neural networks can be trained to detect abnormally that do not fit a pattern because they are good at remembering patterns, for example, learning the daily routine of a person in order to send alert one when something goes wrong, after learning the patterns of such behavior (Shiffman, 2014).

## 4.0 ARTIFICIAL INTELLIGENCE (AI)

AI is tied with cognitive and biological sciences and a subfield of computer science concerned with simulation, modelling, and computing techniques investigating intelligent behavior (Sun, 2008). AI can be simply described as a collection of hard problems which can be solved by humans and other living things, but for which the algorithm for solving them is not available (Zhang, 2011). Research into AI builds on our understanding of the brain, its evolutionary development, and provides knowledge on the functioning and working of the brain, and processes of biological evolution. Although a subfield of computer science, AI draws from neuroscience particularly in how the brain enables humans to think much of which has remained an enigma. However, significant ventures in the field of AI have enabled scientists to continue to progress towards understanding the thought processes in the brain. ANNs are an important part of AI that process data and exhibit some intelligent behaviors (Agarwal, 2014). AI can be applied in the following areas namely:

1. Intelligent Agents
2. Information Retrieval
3. Electronic Commerce
4. Data Mining
5. Bioinformatics
6. Natural Language
7. Expert Systems

The benefit of AI and neural networks in neurobiology is quite enormous. The understanding of neural networks is used in medicine, psychological science, behavioral analysis and in the treatment of diseases and defects of the nervous system. Here ANNs is incorporated as a research tool in developing an understanding of brain neural networks by simulating the brain. Biological Neural Networks are inherently fault tolerant, for example in frequent cases of partial nervous system or brain damage without disruption of life itself (Mano, 2014). ANNs also exhibit a similarly high level of fault tolerance because of their highly distributed and modular nature, however if one particular component or a group of components fails, certain functions may not be performed (Mano, 2014). Nonetheless, the capabilities of the intact components are retained, and the networks do not completely fail. This makes them seemingly fault tolerant.

Furthermore, AI has been applied to perform intelligent task such as in expert systems used by ford mechanics track down and fix engine problems (Zhang, 2011). Moreover, in an airline scheduling program produced with the aid of expert systems offer a graphical user interface to help solve complex airport scheduling problems because it can show graphical images of approaching and circulating planes at the airport, and concourses with planes at their gate (Zhang, 2011). The two major research areas in AI are:

**(a) Artificial Neural Networks:** This area of research involves creating a model that mimics the brain and training it for pattern recognition.
**(b) Genetic Algorithms:** This area of research deals with evolving solutions to complex problems that is hard to control using other methods. These algorithms mimic mechanisms of genetics and natural selection that evolve according to the survival of the fittest leading to the generation of a new set of strings (Roetzel et al. 2020).

### *Benefits of ANNs in AI*

ANNs are very much involved in so many exciting applications of AI. Their benefit includes:

1. Backpropagation Nets
Backpropagation nets learn to generalize and classify patterns. When they are presented with a pattern, the interconnections between the artificial neurons are adjusted until they give a correct response. Backpropagation nets are one of the very popular ANNs; their foundational configuration have to do with interconnected layers of neurons (Mano, 2014). The patterns in this network topology cause unidirectional information flow, then the errors "backpropagate" in changing the weights of connections between layers in another direction (Mano, 2014). A successful example of backpropagation nets is NetTalk which was invented by Terry Sejnowski at the Salk Institute in La Jolla, California (Mano, 2014). This net learns to read English or any other language and is widely used to read to blind people. Basically, after sufficient training of backpropagation nets with a number of patterns, they give correct responses to patterns they have never seen (Mano, 2014).

2. Hopfield Nets

The Hopfield nets was invented by John Hopfield, a physicist at California Institute of Technology (Patel et al. 2014). The basic configuration of the Hopfield net is that every artificial neuron is interconnected to other neurons. These nets memorize collections of patterns; when given a section of the patterns or a distorted pattern, the net delivers the complete pattern (Mano, 2014). For instance, given a partial print or a smudged print, the Hopfield net can deliver the complete fingerprint (Patel, 2014). Hopfield nets have been applied in fingerprint recognition. NASA uses Hopfield nets to orient deep-space craft by visual star fields. When the craft looks at a picture of the stars, a Hopfield net can match the view with the known pictures of the stars to orient the craft (Mano, 2014).

3. Self-Organizing Maps

This was invented by Teuvo Kohonen a Finnish professor and is simply known as Kohonen nets. The basic configuration consists of each artificial neuron connected to only its neighbors (Rodriguez et al 2015). Kohonen nets reduce the complexity of experimentally obtained data because repeated training of the Kohonen net with an n-dimensional data set can produce a lower dimensional data set that captures the essential nature of the n-dimensional data set in a much simpler form (Rodriguez et al 2015). A major application of self-organizing maps is in the implementation in several projects looking for a simpler way to understand the Internet. Kohonen nets are regularly used as a preprocessor for other types of ANN (Mano, 2014).

## Neural networks, conventional computers and the human brain

Research into the use, benefits and applications of neural networks cover a wide range of topics ranging from neurobiology to statistical physics and machine learning; it encompasses so many fields such as neuroscience, computer science, engineering, statistics, cognitive science, physics, biology and philosophy. Scientists from these disciplines are concerned about the exciting and complex nature of neural network. For example, a computer scientist wants to find more about learning systems and neural nets studying properties of non-symbolic processing of information, while statisticians use them in classification models, and nonlinear regression (Heskes and Barber, 2014). The capabilities of neural networks are also used by engineers in automatic control and signal processing, while consciousness and models of thoughts are studied by cognitive scientists using knowledge of neural networks (Heskes and Barber, 2014). In neuroscience, neural networks are used to model brain function while physicists employ neural networks to model statistical mechanics, and biologists use them to understand sequences of nucleotide (Heskes and Barber, 2014).

The brain's network of neurons contrasts with conventional computers because it forms a massively parallel processing system unlike the execution of a single series of instructions in conventional computers (Willamette 2014). The similarities and contrast between the brain and the computer is based on the following:

**(a) Processing Element:** While the brain has $10^{14}$ synapses, a computer has $10^{18}$ transistors as their processing element.

**(b) Processing Speed:** the human brain is composed of about 100 billion neurons (Herculano-Houzel, 2009) while a computer has less than 1 million processors (Zhang, 2011). Also, while the human brain has a processing speed of 100 Hz, that of a computer is $10^9$ Hz.

**(c) Style of Computation:** In the human brain, massive execution of computational processes such as visual perception occurring within a timeline less than 100ms in the brain (represents 10 processing steps) unlike in serial centralized computations (Zhang, 2011).

**(d) Fault Tolerance:** The human brain is fault tolerant, which means that a partial recovery could be possible after damage to areas of the brain if other healthy units can learn the functions of the damaged parts; however, this is not true with computers (Patel, 2014).

**(e) Intelligence and Consciousness:** human self-awareness and intelligence is supported by the brain but conventional computers have not yet been able to do this.

**(f) Learning**: The human brain can learn to reorganize itself from experience unlike in conventional computer, where little learning occurs (Zhang, 2011).

Biological neural networks use a different problem solving approach than algorithms used by current computers. To find solution for a problem computers follow an algorithmic approach that finds the task difficult if specific steps are unknown and this limits its problem-solving capability (Stergiou and Siganos, 2014). On the other hand, neural networks learn by example, so they cannot be programmed to perform a specific task (Stergiou and Siganos, 2014). The selection process must be done carefully otherwise useful time is wasted or the network might function incorrectly.

A major disadvantage of neural networks is that its operation can be unpredictable (Stergiou and Siganos, 2014). In conventional computers, a cognitive approach is implemented where the problem is known and the instructions are set and converted to high-level program language understood by the computer. Also, software and hardware errors are common and inevitable. Despite these differences, conventional algorithmic computers and neural networks complement each other; while there are some tasks more suited to a computational operation such as arithmetic's, task such as facial recognitions are more suitable for neural networks (Stergiou and Siganos, 2014).

### Why we need brain-like intelligence

Scientists have spent a lot of time researching and implementing complex solutions involving brain-like intelligence. It is obvious that some task very easily completed by a computer, but tasking for humans. If a computer is to find the square root of 864,900 for example, an algorithm implemented using a line of code produces the output in less than a millisecond; but for humans this will prove a difficult task and will require more time (Sawicki, 2014). On the other hand, some task that are simple for humans, are not so easy for a computer. If a human is shown a picture of a mice or an African giant rat, they will be able to identify it as they do not need a machine to perform this task.

One of the reasons why we need brain like intelligence is to perform task challenging for a machine but easy for humans. An example is pattern recognition used in computing. Applications that require brain-like intelligence includes a collection such as facial recognition, optical character recognition like converting handwritten or printed scans into digital text. These neural network applications use AI algorithms. Another reason is that a computerized neural network performs optimally than the brain in the context of speed. The brain cannot process information in the amount of time it takes a computer due to the inability to concentrate for a very long period to perform a task (Sawicki, 2014). More also, the human brain programs itself in response to input from the its senses and when there are large number of input variables, the task becomes very difficult to visualize, although the brain has the potential to eventually solve this kind of problem (Sawicki, 2014).

Naturally unassuming, it is extremely difficult for a human to solve tasks involving a twelve-dimensional boundary divided by a thirteen-dimensional space into likelihoods of events because our brains learned from infancy how to perform three-dimensional tasks so anything more would be difficult to learn. While conventional computers and the brain share some similarities and differences, in order to achieve true human-level intelligence a brain-like information processing is needed (Zhang, 2011).

## 5. CONCLUSION

It is very clear that neural networks are beneficial for biology, medicine and computing because they can be trained to learn from example which makes them relevant in providing solution for a catalog of modern day problems. ANNs are computationally useful for real time applications because of their quick response and parallel architecture. In biology and medicine ANNs have been used in biological systems modelling of living organisms to examine their internal architecture. Even though the reasoning foundations of neural networks have a huge potential, scientist will achieve optimal output by combined them with computing. In general, putting a neural network into a computer allows it to make intelligent judgments. Computerized neural networks can function in ways that represents an improvement upon the brain's own processing mechanisms. Although it is easy to visualize how the huge number of neurons in the brain could provide the computational power, it is difficult to explain why the brain is not better at forming complex mathematical judgments.

In all, neural network is a rich area of research which has the potential to simulate a greater range of the operations of the brain than with only computational models. It should be noted that neural network does not perform magic, but can produce very exciting results if used intelligently, as this paper has attempted to explain some of its benefits, and the kind of task that a neural network excels at in computation. The benefit of understanding and applying neural networks in medical science is quite enormous; it is therefore recommended that computational studies involving the use of artificial neural networks should be incorporated in the field of neuroscience. More also, research in the emerging field of computational neuroscience should be highly encouraged by setting up facilities and training personnel in the field.

## Acknowledgment

The content of this manuscript has been published [in part] as part of the Seminar of Nwadiugwu C. Martin at the University of Ilorin, Nigeria

**COMPLIANCE WITH ETHICAL STATEMENTS**

(a) Funding: The author declares no research funding for this review.
(b) Conflict of Interest: The author declares no conflict of interest
(c) Availability of Data and Material: Materials used are duly referenced.
(d) Authors' Contribution: Martin Nwadiugwu: Developed the study question and typed the manuscript.

# ABOUT THE AUTHOR

*Nwadiugwu Chuks Martin:* Department of Bioinformatics, University of Nebraska; Department of Anatomy, University of Ilorin, Ilorin, Nigeria.